\documentclass[12pt]{article}
\usepackage[margin=25mm]{geometry}
\usepackage{amsmath,mathtools,amsfonts,bbm}
\usepackage{graphicx,psfrag,epsf}
\usepackage{enumerate}
\usepackage{natbib}
\usepackage{amssymb}
\usepackage{epstopdf}
\usepackage{enumerate}
\usepackage{placeins}
\usepackage{setspace}
\usepackage{float}
\usepackage{soul,color}
\usepackage{placeins}
\usepackage{upgreek}
\usepackage{algorithm,algpseudocode,tabularx}
\usepackage{caption}
\usepackage{subcaption}
\usepackage{enumitem}
\usepackage{multirow}
\usepackage{amssymb,graphics,amsthm, comment}
\usepackage{url}
\RequirePackage[colorlinks,breaklinks,linkcolor=blue,citecolor=blue,urlcolor=blue]{hyperref}
\usepackage{multicol}
\usepackage{listings} 
\usepackage{xcolor} 
\immediate\write18{texcount -tex -sum  \jobname.tex > \jobname.wordcount.tex}

\graphicspath{{./plots/}}
\newtheorem{theorem}{Theorem}[section]

\theoremstyle{remark}

\theoremstyle{plain}
\theoremstyle{definition}
\newtheorem{remark}{{\bf Remark}}
\newtheorem{assumption}{{\bf Assumption}}

\DeclareMathOperator*{\argmin}{arg\,min}
\DeclareMathOperator*{\argmax}{arg\,max}

\DeclarePairedDelimiter\floor{\lfloor}{\rfloor}

\newcommand{\Rom}[1]{\MakeUppercase{\romannumeral #1}}
\newcommand{\rom}[1]{\romannumeral #1}
\newcommand{\trans}{\mathrm{T}}
\newcommand{\mbf}[1]{#1}

\allowdisplaybreaks

\providecommand{\keywords}[1]
{
  \small	
  \textbf{\textit{Keywords---}} #1
}

\title{Varying coefficient model for longitudinal data with informative observation times}
\author{Yu Gu$^{1}$, Yangjianchen Xu $^{2}$,  and Peijun Sang$^{2}$ \\
	    \small $^{1}$Department of Statistics and Actuarial Science, University of Hong Kong \\
        \small $^{2}$Department of Statistics and Actuarial Science, University of Waterloo \\
}
\date{} 

\begin{document}
\maketitle

\begin{abstract}
Varying coefficient models are widely used to characterize dynamic associations between longitudinal outcomes and covariates. Existing work on varying coefficient models, however, all assumes that observation times are independent of the longitudinal outcomes, which is often violated in real-world studies with outcome-driven or otherwise informative visit schedules. Such informative observation times can lead to biased estimation and invalid inference using existing methods. In this article, we develop estimation and inference procedures for varying coefficient models that account for informative observation times. We model the observation time process as a general counting process under a proportional intensity model, with time-varying covariates summarizing the observed history. To address potential bias, we incorporate inverse intensity weighting into a sieve estimation framework, yielding closed-form coefficient function estimators via weighted least squares. We establish consistency, convergence rates, and asymptotic normality of the proposed estimators, and construct pointwise confidence intervals for the coefficient functions. Extensive simulation studies demonstrate that the proposed weighted method substantially outperforms the conventional unweighted method when observation times are informative. Finally, we provide an application of our method to the Alzheimer's Disease Neuroimaging Initiative study.
\end{abstract} \hspace{10pt}


\section{Introduction}\label{sec-intro}

\begin{figure}[!ht]
	\centering
   {\includegraphics[width=14cm]{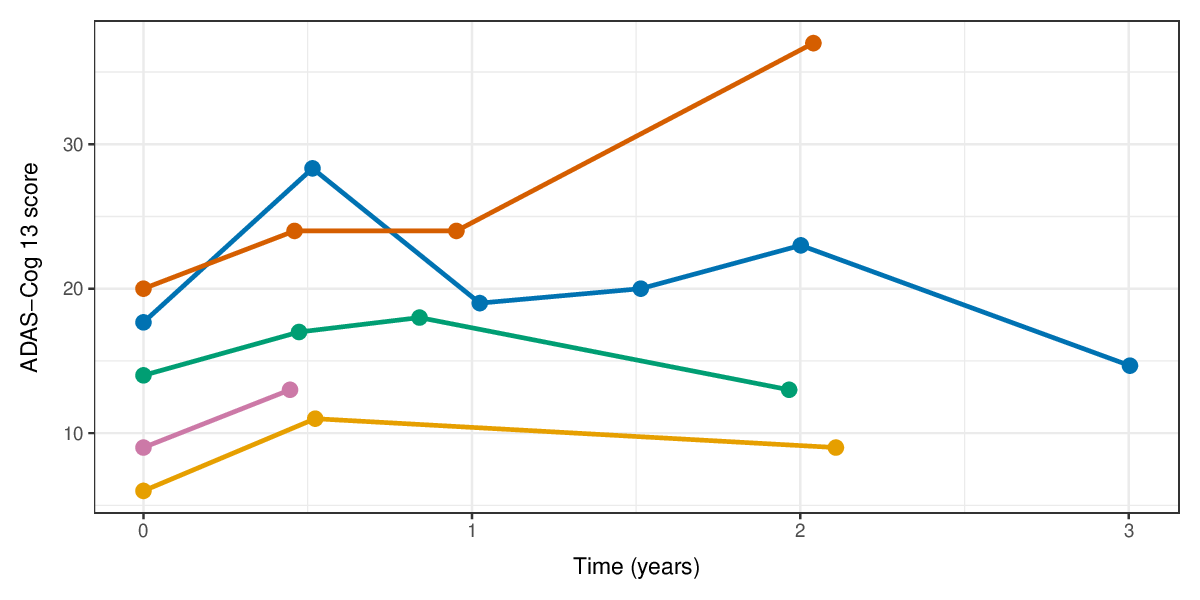}}
	\caption{Observed ADAS-Cog 13 trajectories for five randomly selected subjects from the ADNI study.}
	\label{fig:traj_adas13}
\end{figure}

Modeling the dynamic relationships between a longitudinal outcome and covariates is a fundamental task in biomedical research. In the study of Alzheimer's disease, for example, a primary objective is to characterize the effects of covariates, such as age, gender, years of education, and the number of ApoE-$\epsilon$4 alleles, on the progression of cognitive impairment, which is typically measured by the ADAS-Cog 13 scores at a sequence of observation times. However, this task is complicated by two major challenges. First, the effects of these covariates typically exhibit time-varying patterns as the disease progresses \citep{Jack2013}, requiring flexible modeling approach that allow these effects to vary smoothly over time. Second, the observation times are often informative in the sense that they are stochastically dependent on the longitudinal outcome \citep{pullenayegum2016longitudinal}. In the Alzheimer’s Disease Neuroimaging Initiative (ADNI) \citep{petersen2010}, although clinical visits are pre-scheduled for the subjects, their actual attendance is voluntary. As illustrated in Figure~\ref{fig:traj_adas13}, subjects with higher ADAS-Cog 13 scores tend to return for clinical visits more frequently, indicating dependence between the outcome and the observation time processes. Ignoring either time-varying covariate effects or informative observation times may lead to biased estimation, invalid inference, and misleading analysis results.

The varying coefficient model \citep{Hoover1998} and its extensions have been extensively used to estimate time-varying covariate effects due to their flexibility and interpretability. Specifically, \cite{Huang2002} and \citet{huang2004polynomial} refined the estimation procedure of \citet{Hoover1998} through basis function approximations and established the consistency and asymptotic normality of the coefficient estimators. \cite{fan2007analysis} studied a semiparametric varying coefficient partially linear model but focused on the estimation of the covariance function of the random error process. \cite{zhu2012multivariate} considered a varying coefficient model with multivariate longitudinal outcomes and established the asymptotic properties of the coefficient estimators. \cite{zhang2013time} and \cite{zhang2015varying} extended the model of \cite{Hoover1998} by allowing for nonlinear relationships between the outcome and covariates. All the aforementioned work assumes that the longitudinal outcomes are independent of the observation times and thus may yield inconsistent estimators when the observation times are informative.

However, existing methods for handling informative observation times have primarily focused on estimation of the mean function of the outcome process, rather than the time-varying covariate effects. \cite{lin2004analysis} proposed an inverse intensity weighting approach to account for informative observation times and constructed estimating equations for the conditional mean function of the outcome process. More recent work has focused on marginal mean estimation: \cite{weaver2023functional} introduced a latent factor model to account for the dependence between the outcome and observation time processes; \cite{xu2024bias} modeled this dependence using two latent Gaussian processes and derived a bias-corrected estimator for the mean function; and \cite{sang2025functional} adopted the inverse intensity weighting and estimated the mean function using penalized splines. These methods cannot be directly applied to varying coefficient models, as a proper method requires correction for informative observation times at the level of the covariate-outcome relationship, which fundamentally alters both the numerical implementation and the theoretical analysis. To the best of our knowledge, no existing methods address informative observation times for the varying coefficient model.

In this article, we fill this gap by developing estimation and inference procedures for the varying coefficient model with informative observation times. We allow the intensity function of the observation time process to depend on a subject’s entire observed history, thereby accounting for dependence between the outcome and observation time processes. We establish the identification of the marginal mean function of the outcome process through inverse intensity weighting. This identification result enables sieve estimation of the time-varying coefficients by solving a weighted least squares problem, yielding closed-form estimators that are computationally efficient. We further establish the consistency, convergence rates, and asymptotic normality of the proposed estimators, which facilitate the construction of pointwise confidence intervals for the coefficients. As far as we are concerned, the asymptotic distribution of the estimated slope function has not been established in vary coefficient models with informative observation times. In particular, \cite{sang2025functional} only established consistency and convergence rates for the estimated mean and covariance functions of the longitudinal outcome. Therefore, they did not develop a statistically sound tool to quantify the estimation uncertainty.
We evaluate the finite-sample performance of the proposed estimation and inference procedures through extensive simulation studies. Finally, we describe an application to the ADNI study in which the effects of selected biomarkers on the progression of cognitive impairment are estimated.

	The remainder of the article is organized as follows. Section \ref{sec:preliminary_setup} introduces the data structure, the varying coefficient model for the outcome process, and the intensity function for the observation time process. In Section \ref{sec:estimation}, we develop a sieve estimation approach that incorporates inverse intensity weighting. Section \ref{sec:theory} establishes the asymptotic properties of the proposed estimators and presents the construction of pointwise confidence intervals for the coefficients. Section \ref{sec:simulation_studies} reports results from extensive simulation studies evaluating the performance of the proposed estimators and confidence intervals. In Section \ref{sec:real_data_application}, we analyze the data from the ADNI study. Section \ref{sec:conclusion} concludes the article. Technical proofs and additional numerical results are provided in the Supplementary Material.

\section{Preliminary setup} \label{sec:preliminary_setup}
\subsection{Data and outcome model} \label{subsec:data_and_model}
Consider a longitudinal study of $n$ independent subjects. 
For the $i$th subject ($i=1,\dots,n$), let $Y_i(t)$ denote the outcome process, $\mbf{X}_i(t) = \{X_{i1}(t), \dots, X_{id}(t)\}^{\trans}$ denote a $d$-vector of potentially time-dependent covariates, and $C_i$ denote the follow-up time. Both the outcome and covariate processes are observed only at a finite sequence of time points $0\leq t_{i1}<\cdots<t_{im_{i}}\leq C_{i}$, where $m_{i}$ is the total number of observations for the $i$th subject. These observation times can be represented by a counting process $N_i(t) = \sum_{j = 1}^{m_i} \mathbbm{1}(t_{ij} \leq t)$, for $t \in [0, C_i]$, where $\mathbbm{1}(\cdot)$ is the indicator function. 
We use an overline to denote the history of a stochastic process, e.g., $\overline{N}_i(t) = \{N_i(s): s\in [0,t]\}$ is the history of the counting process $N_i(\cdot)$ up to time $t$; and we use an overline together with a superscript ``$obs$'' to denote the observed history of the outcome and covariate processes, i.e.,
\begin{gather*}
	\overline{Y}_i^{obs}(t) = \left\{Y_i(s):\; dN_i(s)=1, s\in[0,t]\right\}, \\
	\overline{\mbf{X}}_i^{obs}(t) = \left\{\mbf{X}_i(s):\; dN_i(s)=1, s\in[0,t]\right\},
\end{gather*}
where $dN_i(t)\in\{0,1\}$ is the jump indicator of $N_{i}(\cdot)$ over the infinitesimal interval $[t, t + dt)$.
The observed data consist of independent observations $\overline{\mathcal{O}}_i(t)=\{\overline{N}_i(t), \overline{Y}_i^{obs}(t), \overline{\mbf{X}}_i^{obs}(t)\}$ ($i=1,\ldots,n$), where the $i$th subject is observed up to time $C_i$.

Suppose that the study period is $[0,\tau]$, where $\tau$ is the upper bound of the support of $C_i$.
We relate the outcome process $Y_i(t)$ to the covariate process $\mbf{X}_i(t)$ through the varying coefficient model:
\begin{equation} \label{varying_coefficient_model}
	Y_i(t) = \mbf{\beta}(t)^{\trans}\mbf{X}_i(t)+b_i(t)+\epsilon_i(t), \quad \text{ for } t\in[0,\tau],
\end{equation}
where $\mbf{\beta}(t) = \{\beta_1(t), \dots, \beta_d(t)\}^{\trans}$ is a $d$-vector of unknown time-varying coefficients, $b_i(t)$ is a latent subject-specific stochastic process accounting for the within-subject dependence, and $\epsilon_i(t)$ is the measurement error process. 
We assume that $\mbf{X}_i(t)$ contains the constant 1, $b_i(t)$ are zero-mean stochastic processes with covariance function $\Sigma_b(s,t)$, $\epsilon_i(t)$ are zero-mean stochastic processes with covariance function $\sigma_{\epsilon}^2(t)\mathbbm{1}(s=t)$, and $\mbf{X}_i(t)$, $b_i(t)$, and $\epsilon_i(t)$ are mutually independent.

Model \eqref{varying_coefficient_model} and its variants have been extensively studied in the literature. In existing work, the observation times for the outcome process can be either densely spaced \citep{zhu2012multivariate} or sparsely spaced \citep{fan2007analysis, zhang2013time}, and are assumed to be independent of the outcome process. The independence assumption is widely adopted because it substantially simplifies both the estimation procedure and theoretical development. However, as illustrated by \cite{weaver2023functional}, \cite{xu2024bias}, and \cite{sang2025functional}, the assumption is often violated in practice and thus leads to biased estimation of the mean function of the outcome process. In this article, we focus on the estimation and inference for the time-varying coefficients of model~\eqref{varying_coefficient_model} when the observation times are sparsely spaced and informative.

\subsection{Inverse intensity weighting}
\label{sub:inverse_intensity_weighting}
We model the dependence between the outcome and observation time processes through the conditional intensity function of the counting process $N_i(t)$. Let $\mathcal{F}_{it}$ denote the filtration generated by $\{\overline{\mathcal{O}}_i(t-), Y_i(t)\}$ for $i = 1, \ldots, n$ and $t \in [0, \tau]$. We make the following assumption on the counting process $N_i(t)$ and the follow-up time $C_i$.
\begin{assumption} \label{ass:ident}
	The following hold:
	\begin{itemize}
		\item[(\rom{1})] $E\{dN_i(t)|\mathcal{F}_{it}\} = E\{dN_i(t)|\overline{\mathcal{O}}_i(t-)\}$.
		\item[(\rom{2})] The conditional density of $C_i$ at time $t$ given $\{Y_i(\cdot), \mbf{X}_i(\cdot), b_i(\cdot)\}$ depends only on $\overline{\mbf{X}}_i(t)$ and is noninformative about the unknown parameters in model \eqref{varying_coefficient_model}.
	\end{itemize}
\end{assumption}
Let $\lambda_i\{t|\overline{\mathcal{O}}_i(t-)\}$ denote the conditional intensity function of $N_i(t)$ given $\overline{\mathcal{O}}_i(t-)$, i.e., $E\{dN_i(t)|\overline{\mathcal{O}}_i(t-)\} = \lambda_i\{t|\overline{\mathcal{O}}_i(t-)\}dt$.
Assumption \ref{ass:ident}(i) indicates that the observation time process only depends on the past observed history, not on the current outcome. 
Assumption \ref{ass:ident}(ii) is an independent censoring condition for counting processes, which is standard in the survival analysis literature (e.g., \citealt{zeng2005asymptotic}).
Under Assumption \ref{ass:ident}(i), we have 
\begin{equation}  \label{eq:mean-identify}
	\lambda_i\{t|\overline{\mathcal{O}}_i(t-)\}^{-1}E\{Y_i(t)dN_i(t)\} 
	= \lambda_i\{t|\overline{\mathcal{O}}_i(t-)\}^{-1}E\left[E\{Y_i(t)dN_i(t)|\mathcal{F}_{it}\}\right] 
	= E\{Y(t)\}dt.
\end{equation}
Thus, weighting by $\lambda\{t|\overline{\mathcal{O}}(t-)\}^{-1}$ creates a pseudo-population in which the observation times are no longer associated with the outcome process $Y(t)$. In this pseudo-population, valid estimation and inference can be conducted as if all the observed outcomes were sampled at random, i.e., independently of the observation times. 

We specify that the conditional intensity function $\lambda_i\{t|\overline{\mathcal{O}}_i(t-)\}$ satisfies the proportional intensity model:
\begin{equation} \label{PI_model}
	\lambda_i\{t|\overline{\mathcal{O}}_i(t-)\} = \lambda_0(t)\exp[\mbf{\gamma}^{\trans}\mbf{g}\{\overline{\mathcal{O}}_i(t-)\}],
\end{equation}
where $\lambda_0(t)$ is an arbitrary baseline intensity function, $\mbf{\gamma}$ is a vector of unknown regression parameters, and $\mbf{g}(\cdot)$ is a vector of prespecified functions of $\overline{\mathcal{O}}_i(t-)$.
One typical choice for $\mbf{g}\{\overline{\mathcal{O}}_i(t-)\}$ is to use the most recently observed outcome and covariate values. The semiparametric form of model \eqref{PI_model} offers substantial flexibility by leaving the baseline intensity function completely unspecified, thereby avoiding restrictive parametric assumptions on the observation time process. In addition, this specification induces a first-order Markov structure for the observation time process, in the sense that the effect of past history on the current intensity operates entirely through these last observed values. Such a structure is natural and often reasonable, because in practice the most recent measurements typically represent the information available to guide scheduling decisions.

To account for the informative observation times, as implied by \eqref{eq:mean-identify}, it suffices to assign the following inverse-intensity weight to each observation:
\begin{equation} \label{weights}
	w_{ij} = \lambda_0(t_{ij})^{-1}\exp[-\mbf{\gamma}^{\trans}\mbf{g}\{\overline{\mathcal{O}}_i(t_{ij}-)\}],
\end{equation}
for $i=1,\dots,n$ and $j=1,\dots,m_i$.
As shown in the following sections, the weighted data can be analyzed as if they were collected at non-informative observation times, enabling the use of standard estimation and inference procedures for varying coefficient models.

\section{Estimation procedure} \label{sec:estimation}
We employ a two-stage estimation procedure. 
In the first stage, we estimate the inverse-intensity weights $w_{ij}$ in \eqref{weights} using maximum partial likelihood and kernel smoothing. 
In the second stage, we incorporate these estimated weights into the estimation of the varying coefficient model~\eqref{varying_coefficient_model}. 

\subsection{Estimation of weights} \label{sub:estimation_of_weights}
We first estimate $\mbf{\gamma}$ in model~\eqref{PI_model} by maximizing the log partial likelihood:
\[
\widehat{\mbf{\gamma}} = \argmax_{\mbf{\gamma}} \sum_{i=1}^n \int_0^{C_i} \left[\mbf{\gamma}^{\trans}\mbf{g}\{\overline{\mathcal{O}}_i(t-)\}-\log S^{(0)}(t;\mbf{\gamma})\right] dN_i(t),
\]
where $S^{(0)}(t;\mbf{\gamma}) = \sum_{i=1}^n \mathbbm{1}(C_i\ge t)\exp[\mbf{\gamma}^{\trans}\mbf{g}\{\overline{\mathcal{O}}_i(t-)\}]$.
The above maximization can be solved using the Newton--Raphson algorithm. 
Let $\Lambda_0(t)=\int_0^t \lambda_0(s)ds$ be the cumulative baseline intensity function. 
We then estimate $\Lambda_0(t)$ by the Breslow--Aalen-type estimator \citep{Breslow,Aalen1978}
\[
\widehat{\Lambda}_0(t) = \int_0^t\frac{\sum_{i=1}^n \mathbbm{1}(C_i\ge u)dN_i(u)}{S^{(0)}(u;\widehat{\mbf{\gamma}})}.
\]
Finally, we obtain a kernel-smoothed estimator for $\lambda_0(t)$ by
\begin{equation} \label{eq:estbaseline}
	\widehat{\lambda}_0(t) = h_{n}^{-1}\int_0^{\tau}K\left(\frac{t-s}{h_{n}}\right)d\widehat{\Lambda}_0(s),
\end{equation}
where $K(\cdot)$ is the kernel function and $h_{n}$ is the bandwidth of order $n^{-1/5}$ as suggested in Chapter \Rom{4}.2 of \citet{andersen1993statistical}. 
Following \citet{muller1994hazard}, we use a data-adaptive bandwidth of the form $h_{n} = c\tau N^{-1/5}$, where $c>0$ is a tuning parameter and $N = \sum_{i=1}^n m_i$. 
Based on our numerical experience, the estimation of $\lambda_0(t)$ is not sensitive to the specific choice of $K(u)$ or $h_n$; the results differ only in the third decimal place when alternative kernel functions are used or when $c$ ranges from 0.1 to 1.0 in the bandwidth specification.
Thus, in all subsequent numerical studies, we report results using the Epanechnikov kernel $K(u)=0.75(1-u^2)\mathbbm{1}(-1\le u\le 1)$ with bandwidth $h_{n} = 0.1\tau N^{-1/5}$.
The inverse-intensity weight assigned to the $j$th observation of the $i$th subject can then be estimated by 
\begin{equation} \label{eq:estweight}
	\widehat{w}_{ij} = \widehat{\lambda}_0(t_{ij})^{-1}\exp[-\widehat{\mbf{\gamma}}^{\trans}\mbf{g}\{\overline{\mathcal{O}}_i(t_{ij}-)\}].
\end{equation}

\subsection{Estimation of varying coefficients} \label{sub:estimation_of_varying_coefficients}
Now we estimate the time-varying coefficient $\mbf{\beta}(t)$ in model~\eqref{varying_coefficient_model} using weighted least squares with the weights in \eqref{eq:estweight}.
Since each component of $\mbf{\beta}(t)$ is unknown and nonparametric, 
we employ the idea of regression splines and approximate $\mbf{\beta}(t)$ by a sieve of B-spline functions. Specifically, let $0 = \kappa_0 < \kappa_1 < \cdots < \kappa_{q_n - z} < \kappa_{q_n - z + 1} = \tau$ denote a sequence of knots that partition $[0, \tau]$ into $q_n - z + 1$ subintervals of equal length.  Let $\mbf{B}(t) = \{B_1(t), \ldots, B_{q_n}(t)\}^{\trans}$ denote a vector of $q_n$ normalized B-spline basis functions that satisfy the following properties: (i) each of them is a polynomial of degree up to $z - 1$ in each subinterval and (ii) their $(z - 2)$th order derivatives are continuous at the interior knots $\kappa_1, \dots, \kappa_{q_n - z}$. Under certain smoothness conditions on $\mbf{\beta}(t)$ that will be specified in $\mathsection$\ref{sec:theory}, we can accurately approximate
$\mbf{\beta}(t)$ by $\mbf{A}^{\trans}\mbf{B}(t)$, where $\mbf{A}$ is a $q_n\times d$ matrix of unknown spline coefficients.

Let $Y_{ij}$ and $\mbf{X}_{ij}$ denote the outcome and covariates observed at time $t_{ij}$, respectively, for $i=1,\dots,n$ and $j=1,\dots,m_i$.
Also write $\mbf{B}_{ij} = \mbf{B}(t_{ij})$.
We estimate the matrix $\mbf{A}$ by the following weighted least squares estimator:
\begin{equation} \label{eq:estslope}
	\widehat{\mbf{A}} = \argmin_{\mbf{A}} \sum_{i=1}^n\sum_{j=1}^{m_i} \widehat{w}_{ij} \left(Y_{ij}-\mbf{B}_{ij}^{\trans}\mbf{A}\mbf{X}_{ij}\right)^2.
\end{equation}
Let $\text{vec}(\cdot)$ denote the vectorization of a matrix, $\otimes$ the Kronecker product and $\mbf{a}^{\otimes2} = \mbf{a}\mbf{a}^{\trans}$ for a column vector $\mbf{a}$.
Solving \eqref{eq:estslope} yields a closed-form estimator for the matrix $\mbf{A}$:
\[
\text{vec}(\widehat{\mbf{A}})= \left\{\sum_{i=1}^n\sum_{j=1}^{m_i}\widehat{w}_{ij}\left(\mbf{X}_{ij}\otimes \mbf{B}_{ij}\right)^{\otimes2}\right\}^{-1} \left\{\sum_{i=1}^n\sum_{j=1}^{m_i}\widehat{w}_{ij}\left(\mbf{X}_{ij}\otimes \mbf{B}_{ij}\right)Y_{ij}\right\}.
\]
The estimator for $\mbf{\beta}(t)$ is then given by $\widehat{\mbf{\beta}}(t) = \widehat{\mbf{A}}^{\trans}\mbf{B}(t)$. Consequently, the estimator for the $j$th component of $\mbf{\beta}(t)$, denoted by $\beta_j(t)$, is the $j$th component of $\widehat{\mbf{\beta}}(t)$, denoted by $\widehat{\beta}_j(t)$.

\subsection{Functional principal component analysis} \label{sub:functional_principal_component_analysis}
We consider functional principal component analysis for the latent subject-specific random functions $b_i(t)$ in model~\eqref{varying_coefficient_model}, which are zero-mean stochastic processes with covariance function $\Sigma_b(s,t)$. 
By Mercer's theorem, $\Sigma_b(s,t)$ admits a spectral decomposition $\Sigma_b(s,t) = \sum_{l=1}^{\infty}\theta_l\phi_l(s)\phi_l(t)$, where $\{\theta_l\}_{l=1}^{\infty}$ is a nonincreasing sequence of nonnegative eigenvalues and $\{\phi_l(t)\}_{l=1}^{\infty}$ are the corresponding orthonormal eigenfunctions.
It follows from the Karhunen–Loève expansion that $b_i(t) = \sum_{l=1}^{\infty}b_{il}\phi_l(t)$, where $\{b_{il}\}_{l=1}^{\infty}$ are the functional principal component scores with $E(b_{il}) = 0$ and $\text{cov}(b_{il}, b_{il'}) = \theta_l \mathbbm{1}(l=l')$. 

From the estimation procedure described in $\mathsection$\ref{sub:estimation_of_weights} and $\mathsection$\ref{sub:estimation_of_varying_coefficients}, we obtain the residuals 
\[
R_{ij} = Y_{ij} - \widehat{\mbf{\beta}}(t_{ij})^{\trans}\mbf{X}_{ij} = Y_{ij}-\mbf{B}_{ij}^{\trans}\widehat{\mbf{A}}\mbf{X}_{ij},
\]
for $i=1,\dots,n$ and $j=1,\dots,m_i$.
These residuals can be viewed as noisy observations of the random function $b_i(t)$ with independent measurement errors.
Thus, based on $R_{ij}$'s, we can estimate the covariance function $\Sigma_b(s,t)$, as well as its eigenfunctions $\phi_l(t)$ and eigenvalues $\theta_l$. Since $R_{ij}$'s are sparsely observed, we employ the  method developed in 
\citet{yao2005functional}, which essentially aggregates observations from all subjects to estimate the population-level covariance function $\Sigma_b$. Specifically, define the raw covariance $\widehat{\Sigma}_b(t_{ij}, t_{ij^{\prime}}) = {R}_{ij} {R}_{ij^{\prime}}$ for $j, j^{\prime} \in \{1, \ldots, m_i\}$ and $j\ne j'$. Then, a local linear method is used to estimate $\Sigma_b(s, t)$, with $\widehat{\Sigma}_b(s, t) = \hat{a}_0$ defined as
$$
(\hat{a}_0, \hat{a}_1, \hat{a}_2) = \argmin_{a_0, a_1, a_2} \sum_{i = 1}^n \sum_{j \neq j^{\prime}} \{{R}_{ij} {R}_{ij^{\prime}} - a_0 - a_1(s - t_{ij}) - a_2(t - t_{ij^{\prime}})\}^2 H \left(\frac{s - t_{ij}}{h_b}\right) H \left(\frac{t - t_{ij^{\prime}}}{h_b}\right),
$$
where $H(\cdot)$ denotes a kernel function and $h_b > 0$ denotes the bandwidth. Solving the eigenequations $\int_0^{\tau} \widehat{\Sigma}_b(s, t) \widehat{\phi}_l(s) ds = \widehat{\theta}_l \widehat{\phi}_l(t)$ yields the estimated eigenvalues $\widehat{\theta}_l$ and corresponding eigenfunctions $\widehat{\phi}_l$.  
The entire procedure, including bandwidth selection, can be implemented through the function \texttt{FPCA} in the R package ``fdapace'' \citep{zhou2024fdapace}.

\section{Asymptotic properties and inference} 
\label{sec:theory}
We first present a set of conditions under which the proposed estimation procedure enjoys valid statistical guarantees, and then establish the convergence rates and asymptotic normality of $\widehat{\mbf{\beta}}(t)$. Based on these theoretical results, we construct pointwise confidence intervals for $\mbf{\beta}(t)$.

\subsection{Regularity conditions and asymptotic properties}
For any $p \in \mathbb{N}^+$, let $C^p ([0, \tau])$ denote the class of functions with continuous $p$th order derivatives over $[0, \tau]$. Let $0 = \kappa_0 < \kappa_1 < \cdots < \kappa_{q_n - z} < \kappa_{q_n - z + 1} = 1$ denote the knots of the B-spline basis functions and $[d] := \{1, \ldots, d\}$. For any function $f \in L^2([0, \tau])$, the class of square integrable functions on $[0, \tau]$, define $\|f\| = \left\{\int_0^{\tau} f^2(t) dt\right\}^{1/2}.$
In addition, for any function $f$ defined on $[0,\tau]$, define the supremum norm $\|f\|_{\infty} = \sup_{t\in[0,\tau]}|f(t)|$.
The following regularity conditions are needed to establish the asymptotic properties for $\widehat{\mbf{\beta}}(t)$, where the subscript $i$ is omitted on random elements for simplicity.
\begin{assumption} \label{ass: continuous}
	Each component of  $\mbf{\beta}$ belongs to $C^p ([0, \tau])$ for some $p \geq 2$, i.e., $\beta_j \in C^p([0, \tau])$ for $j \in [d] $.
\end{assumption}

\begin{assumption} \label{ass: spline}
	The order of the spline functions satisfies $z \geq p$ and there exists some constant $M_{\kappa}$ such that
	$\max_{0 \leq j \leq q_n - z}|\kappa_{j + 1} - \kappa_j| \leq M_{\kappa} \min_{0 \leq j \leq q_n - z}|\kappa_{j + 1} - \kappa_j|$. 
\end{assumption}

\begin{assumption} \label{ass: bounded}
	The random vector $\mbf{X}(t)$ is uniformly bounded over $[0,\tau]$. In addition, the eigenvalues of $E\{\mbf{X}(t)^{\otimes2}\}$ are bounded away from 0 and infinity uniformly over $[0,\tau]$.
\end{assumption}


\begin{assumption} \label{ass: error}
	There exists some constant $\delta > 2$ such that $E\{|X_{j}(t)|^{\delta}\}$, $E\{|b(t)|^{\delta}\}$, and $E\{|\epsilon(t)|^{\delta}\}$ are uniformly bounded over $[0,\tau]$ for all $j\in[d]$, where $X_{j}(t)$ is the $j$th component of $\mbf{X}(t)$.
\end{assumption}

\begin{assumption} \label{ass: intensity}
	In the proportional intensity model~\eqref{PI_model}, 
	the baseline intensity function $\lambda_0(t)$ belongs to $C^{r}([0, \tau]$) for some $r \geq p$ and is strictly positive, and the covariates $\mbf{g}\{\overline{\mathcal{O}}(t-)\}$ is almost surely bounded over $[0, \tau]$. 
\end{assumption}

Assumptions \ref{ass: continuous} and \ref{ass: spline} are commonly adopted in the literature of nonparametric smoothing. They ensure that, for any $j \in [d]$, there exists a spline function $\tilde{\beta}_j(t) = \mbf{B}^{\trans}(t) \tilde{\mbf{a}}_j$ such that $\| \beta_j - \tilde{\beta}_j\|_{\infty} = O(q_n^{-p})$; see Lemma 5 of \cite{stone1985}. This approximation result justifies the use of B-spline functions to approximate each coefficient function in the proposed estimation procedure. Assumptions \ref{ass: bounded} pertains to the covariate process and is adopted by \cite{Huang2002} and \cite{huang2004polynomial}. Assumption \ref{ass: error} imposes moment conditions on $\mbf{X}(t)$, $b(t)$, and $\epsilon(t)$, which is satisfied when the related stochastic processes are Gaussian processes; similar conditions are adopted by \cite{li2010uniform} and \cite{zhang2016}.

Assumption \ref{ass: intensity} imposes a smoothness condition on the baseline intensity function and is standard in semiparametric modeling for the intensity function of a counting process, such as the \cite{cox1972} model. This assumption ensures that replacing $\lambda_{0}(\cdot)$ with its estimator $\widehat{\lambda}_{0}(\cdot)$ in the estimated inverse-intensity weight in \eqref{eq:estweight} dose not affect the convergence rate of $\widehat{\beta}(t)$. Specifically, it follows from \cite{andersen1993statistical} that $\widehat{\mbf{\gamma}}$ is $\sqrt{n}$-consistent, and $\widehat{\lambda}_0(t)$ is a consistent estimator for $\lambda_0(t)$ at rate $n^{-r/(2r + 1)}$, provided that the bandwidth $h_n$ satisfies $h_n \asymp n^{-1/(2r + 1)}$ and the kernel $K$ is of order $\floor*r$, the greatest integer strictly less than $r$ \cite[p.~5]{tsybakov2009}. Under this assumption, the convergence rate of $\widehat{\lambda}_0(t)$ is no slower than the $L^2$-convergence rate established in Theorem \ref{thm-slopeconsistency}. Consequently, the use of $\widehat{\lambda}_{0}(\cdot)$ has no impact on the convergence rate of the proposed coefficient function estimator.

We state the $L^2$-convergence rate and the asymptotic normality of $\widehat{\mbf{\beta}}(t)$ in Theorems \ref{thm-slopeconsistency} and \ref{thm-slopeasymnorm}, respectively.

\begin{theorem} \label{thm-slopeconsistency}
	Under Assumptions \ref{ass:ident}--\ref{ass: intensity}, the $j$th estimated coefficient function $\widehat{\beta}_j(t)$ satisfies
	\begin{equation} \label{eq-meanconsistency}
		\|\widehat{\beta}_j - \beta_j\| =  O_p\left\{q_n^{-p} + \left(\frac{q_n}{n}\right)^{\frac{1}{2}}\right\}
	\end{equation}
	for all $j \in [d]$, provided that $\log n/n = o(q_n^{-4})$. 
\end{theorem}

\begin{remark} \label{rmk:slopeest}
	If $q_n \asymp n^{1/(1 + 2p)}$, the $L^2$-convergence rate of $\widehat{\mbf{\beta}}$ is $O_p\{n^{-p/(1 + 2p)}\}$, which is identical to the optimal convergence rate established in \cite{stone1982} for independent and identically distributed data and the rate for varying coefficient additive models for functional data \citep{zhang2015varying}. Once we obtain the convergence rate of $\widehat{\mbf{\beta}}$, we can then employ the techniques developed in \cite{zhou2025theory} to establish the convergence rate for $\widehat{\Sigma}_b$ and the corresponding estimated eigenfunctions $\widehat{\phi}_l$. 
\end{remark}

\begin{theorem} \label{thm-slopeasymnorm}
	Define $\mbf{\Psi}(t)=\mbf{B}(t)\otimes \mbf{I}_{d}$ and $\mbf{Z}_{ij}=\mbf{X}_{ij}\otimes\mbf{B}_{ij}$. Under Assumptions \ref{ass:ident}--\ref{ass: intensity}, $\{n\mbf{\Sigma}_{n}^{-1}(t)\}^{1/2}\{\widehat{\mbf{\beta}}(t)-\mbf{\beta}(t)\}$ converges weakly to a zero-mean multivariate normal random vector with an identity covariance matrix provided that $\log n/n = o(q_n^{-4})$, $q_n^\delta=o(n^{\delta-2})$ and $n=o(q_{n}^{2p+1})$, where $\mbf{\Sigma}_{n}(t)=\mbf{\Psi}(t)^{\mathrm{T}} \mbf{H}_n^{-1} \mbf{\Omega}_n \mbf{H}_n^{-1} \mbf{\Psi}(t)$, $\mbf{H}_{n}=n^{-1}\sum_{i=1}^{n}E(\sum_{j=1}^{m_{i}}w_{ij}\mbf{Z}_{ij}^{\otimes 2})$, and $$\mbf{\Omega}_{n}=n^{-1}\sum_{i=1}^{n}E\left[\sum_{j=1}^{m_{i}}w_{ij}^{2}\mbf{Z}_{ij}^{\otimes 2}\{\Sigma_{b}(t_{ij},t_{ij})+\sigma_{\epsilon}^{2}(t_{ij})\}+\sum_{j\neq k}w_{ij}w_{ik}\mbf{Z}_{ij}\mbf{Z}_{ik}^{\mathrm{T}}\Sigma_{b}(t_{ij},t_{ik})\right].$$
\end{theorem}

\begin{remark} \label{rmk:slopeest-asymp}
	Theorem \ref{thm-slopeasymnorm} enables the statistical inference for $\mbf{\beta}$. In particular, one can construct pointwise confidence intervals for each $\beta_j(t)$ for $t \in [0, \tau]$. Because the analytical form of the asymptotic variance is complex, we resort to the bootstrap procedure to perform inference; details are provided in $\mathsection$\ref{sub:inference_on_varying_coefficients}.
\end{remark}

\begin{remark} \label{rmk:varcomparison}
	\citet{huang2004polynomial} established the asymptotic normality of $\widehat{\mbf{\beta}}(t)$ under the assumption of noninformative observation times. Theorem \ref{thm-slopeasymnorm} extends this result to the setting with informative observation times by incorporating inverse intensity weighting. As shown in $\mathsection$\ref{sec:simulation_studies}, the unweighted estimator of \cite{huang2004polynomial} yields biased coefficient estimates and incorrect variance in the presence of informative observation times, resulting in invalid statistical inference and potentially misleading analysis results. In contrast, the proposed inverse-intensity weighted estimator provides valid inference under informative observation times.
\end{remark}

\subsection{Inference on varying coefficients} \label{sub:inference_on_varying_coefficients}
The estimation of the time-varying coefficients $\mbf{\beta}(t)$ has been discussed in $\mathsection$\ref{sub:estimation_of_varying_coefficients}. We now turn to statistical inference for $\mbf{\beta}(t)$, including variance estimation and the construction of confidence intervals. We estimate the covariance function of $\widehat{\mbf{\beta}}(t)$ using the multiplier bootstrap method introduced in Chapters 2.9 and 3.6 of \citet{vaart1996weak}.
Specifically, at the $l$th bootstrap iteration, we generate random multipliers $\{\xi_{li}\}_{i=1}^n\overset{iid}{\sim} 2\times\text{Bernoulli}(0.5)$ and compute the estimator $\widehat{\mbf{\beta}}^{(l)}(t) = \widehat{\mbf{A}}_l^{\trans}\mbf{B}(t)$, where 
\[
\widehat{\mbf{A}}_l = \argmin_{\mbf{A}} \sum_{i=1}^n\sum_{j=1}^{m_i} \xi_{li} \widehat{w}_{ij} \left(Y_{ij}-\mbf{B}_{ij}^{\trans}\mbf{A}\mbf{X}_{ij}\right)^2.
\]  
We repeat the above procedure $L$ times to estimate the variance of $\widehat{\mbf{\beta}}(t)$.
In particular, for every $t\in[0,\tau]$, the variance of $\widehat{\mbf{\beta}}(t)$ can be estimated by 
\[
\widehat{\mbf{V}}(t) = \frac{1}{L-1}\sum_{l=1}^L \left\{\widehat{\mbf{\beta}}^{(l)}(t)-\overline{\mbf{\beta}}(t)\right\}^{\otimes 2},
\]
where $\overline{\mbf{\beta}}(t) = L^{-1}\sum_{l=1}^L \widehat{\mbf{\beta}}^{(l)}(t)$.
For $j\in[d]$, let $\widehat{V}_{jj}(t)$ denote the $(j,j)$th element of $\widehat{\mbf{V}}(t)$. 
Then, a $(1-\alpha)$ pointwise confidence interval for $\beta_j(t)$ can be constructed by 
\[
\left[\widehat{\beta}_j(t) - z_{1-\alpha/2} \times \sqrt{\widehat{V}_{jj}(t)},\; \widehat{\beta}_j(t) + z_{1-\alpha/2} \times \sqrt{\widehat{V}_{jj}(t)}\right], \quad \text{ for } t\in[0,\tau],
\] 
where $z_{1-\alpha/2}$ is the $(1-\alpha/2)$ quantile of $N(0,1)$.
The numerical results reported in this article are based on $L=100$ bootstrap replicates. Similar strategies to estimate asymptotic variance and construct confidence intervals have been adopted in the literature of functional regression models; see \cite{dette2024statistical} and \cite{xie2023scalable} for example.

\section{Simulation studies} \label{sec:simulation_studies}

In this section, we conduct extensive simulation studies to evaluate the finite-sample performance of the proposed estimation and inference procedures.
\subsection{Practical considerations}
A direct application of regression splines is likely to yield an excessively wiggly fit if too many knots are used. However, using too few knots may lead to an inadequate fit since the spline approximation is not accurate. To address this trade-off,  
\cite{claeskens2009asymptotic} recommended penalized splines. 
Following this approach, we incorporate an additional penalty term $p(\mbf{A};\mbf{\eta})$ into the weighted squared loss function in \eqref{eq:estslope} to estimate $\mbf{A}$. This strategy enables us to
control the smoothness of the estimator for $\mbf{\beta}(t)$ by carefully choosing the tuning parameter vector $\mbf{\eta}= (\eta_1,\dots,\eta_d)^{\trans}$, where each $\eta_{j}>0$. 
One common choice of the penalty is 
\[
p(\mbf{A};\mbf{\eta}) = \frac{1}{2}\text{vec}(\mbf{A})^{\trans}\mbf{S}_{\mbf{\eta}}\text{vec}(\mbf{A}),
\]
where
\[
\mbf{S}_{\mbf{\eta}} = \text{diag}(\eta_1,\dots,\eta_d)\otimes \int_0^{\tau}\left\{\mbf{B}^{(k)}(t)\right\}^{\otimes2}dt,
\]
and $\mbf{B}^{(k)}(t)$ denotes the componentwise $k$th-order derivative of $\mbf{B}(t)$ for some positive integer $k<z$.

In practice, we select the tuning parameters in penalized splines, including $z$, $q_n$, and $\mbf{\eta}$, using generalized cross-validation, while fixing $k=2$ for the matrix $\mbf{S}_{\mbf{\eta}}$. 
We first introduce some notation for aggregation across subjects and observations. 
Recall that $N = \sum_{i=1}^n m_i$ is the total number of observations across subjects.
Let $\mbf{Y} = (Y_{ij}: i=1,\dots,n; j=1,\dots,m_i)^{\trans}$ be the $N$-vector of observed outcomes, and $\mbf{W} = \text{diag}(\widehat{w}_{ij}: i=1,\dots,n; j=1,\dots,m_i)$ be the $N\times N$ diagonal matrix of estimated weights.  
Define $\mbf{Z} = (\mbf{Z}_{ij}^{\trans}: i=1,\dots,n; j=1,\dots,m_i)^{\trans}$, which is an $N\times (dq_n)$ matrix with rows $\mbf{Z}_{ij}^{\trans}$.
By Chapter 3 of \citet{gu2013smoothing} and simple calculation, the generalized cross-validation score for the penalized version of \eqref{eq:estslope} is 
\[
\text{GCV}(z, q_n, \mbf{\eta}) = \frac{N^{-1} \mbf{Y}_w^{\trans}\left(\mbf{I}_N-\mbf{A}_w\right)^2 \mbf{Y}_w}{\left\{N^{-1} \operatorname{tr}\left(\mbf{I}_N-\mbf{A}_w\right)\right\}^2},
\] 
where $\mbf{Y}_w = \mbf{W}^{1/2}\mbf{Y}$, $\mbf{I}_N$ is the $N\times N$ identity matrix, and $\mbf{A}_w = \mbf{W}^{1/2}\mbf{Z}(\mbf{Z}^{\trans}\mbf{W}\mbf{Z}+\mbf{S}_{\mbf{\eta}}/2)^{-1}\mbf{Z}^{\trans}\mbf{W}^{1/2}$.
The optimal tuning parameters are selected by minimizing $\text{GCV}(z, q_n, \mbf{\eta})$.

\subsection{Data generation process}
We generate the outcome process from the following varying coefficient model:
\[ 
Y_i(t) = \beta_1(t)+\beta_2(t)X_{i1}+\beta_3(t)X_{i2}+b_i(t)+\epsilon_i(t), 
\]
where the covariates $(X_{i1}, X_{i2})$ and the random function $b_i(t)$ are generated as follows:
\begin{gather*}
	\begin{pmatrix}
		X_{i1} \\
		X_{i2}
	\end{pmatrix} 
	\sim N\left\{
	\begin{pmatrix}
		0 \\
		0
	\end{pmatrix}, 
	\begin{pmatrix}
		1 & 2^{-0.5} \\
		2^{-0.5} & 1
	\end{pmatrix}
	\right\}, \\
	b_i(t) = b_{i1}\phi_1(t)+b_{i2}\phi_2(t), \\
	b_{i1}\sim N(0, \theta_1), \quad b_{i2}\sim N(0, \theta_2),
\end{gather*} 
and $\epsilon_i(t)$ is a zero-mean Gaussian process with covariance function $0.2 \mathbbm{1}(s=t)$. 
The random elements $(X_{i1}, X_{i2})$, $b_{i1}$, $b_{i2}$, and $\epsilon_i(t)$ are mutually independent. 
We set the study end time to $\tau = 10$ and the parameter values as follows:
\begin{gather*}
	\beta_1(t) = t^2/\tau^2, \quad \beta_2(t) = (\tau-t)^2/\tau^2, \quad \beta_3(t) = 4t(\tau-t)/\tau^2, \\
	\phi_1(t) = \sqrt{2}\sin(2\pi t), \quad \phi_2(t) = \sqrt{2}\cos(2\pi t), \quad \theta_1 = 0.4, \quad \theta_2 = 0.2.
\end{gather*} 
For each subject, we generate the observation times sequentially up to $\tau$ according to the proportional intensity model:
\begin{equation} \label{sim_PI_model}
	\lambda_i\{t|\overline{\mathcal{O}}_i(t-)\} = \exp\left\{Y_i^{obs}(t-) + 0.3X_{i1} + 0.1X_{i2}\right\},
\end{equation}
where $Y_i^{obs}(t-)$ denotes the most recently observed outcome value prior to time $t$.  
We consider sample sizes $n=100$ and $n=200$. 
Under the above simulation setting, the median number of observations is 11 per subject. To better assess the uncertainty of the parameter estimates, we run 200 independent simulation trials for each simulation setting.

\subsection{Simulation results}
We compare the proposed method with two alternatives: the unweighted method and the weighted method using true inverse-intensity weights.
The inference procedures are the same across all methods, except that in \eqref{eq:estslope}, the estimated weights $\widehat{w}_{ij}$ are replaced by 1 for the unweighted method and by the true weights $w_{ij}$ for the weighted method using true weights. 
The evaluation metrics include (integrated) squared errors of the estimated varying coefficients, covariance function, first eigenfunction, and first eigenvalue for $b_i(t)$, defined respectively as $\int_{0}^{\tau} \{\widehat{\beta}_j(t)-\beta_j(t)\}^2dt$ (for $j=1,2,3$), $\int_{0}^{\tau}\int_{0}^{\tau} \{\widehat{\Sigma}_b(s,t)-\Sigma_b(s,t)\}^2dsdt$, $\int_{0}^{\tau} \{\widehat{\phi}_1(t)-\phi_1(t)\}^2dt$, and $(\widehat{\theta}_1-\theta_1)^2$. 
In addition, we evaluate the empirical coverage percentage of the 95\% pointwise confidence intervals for each varying coefficient $\beta_j(t)$. 

\begin{figure}
	\centering
	{\includegraphics[width=14cm]{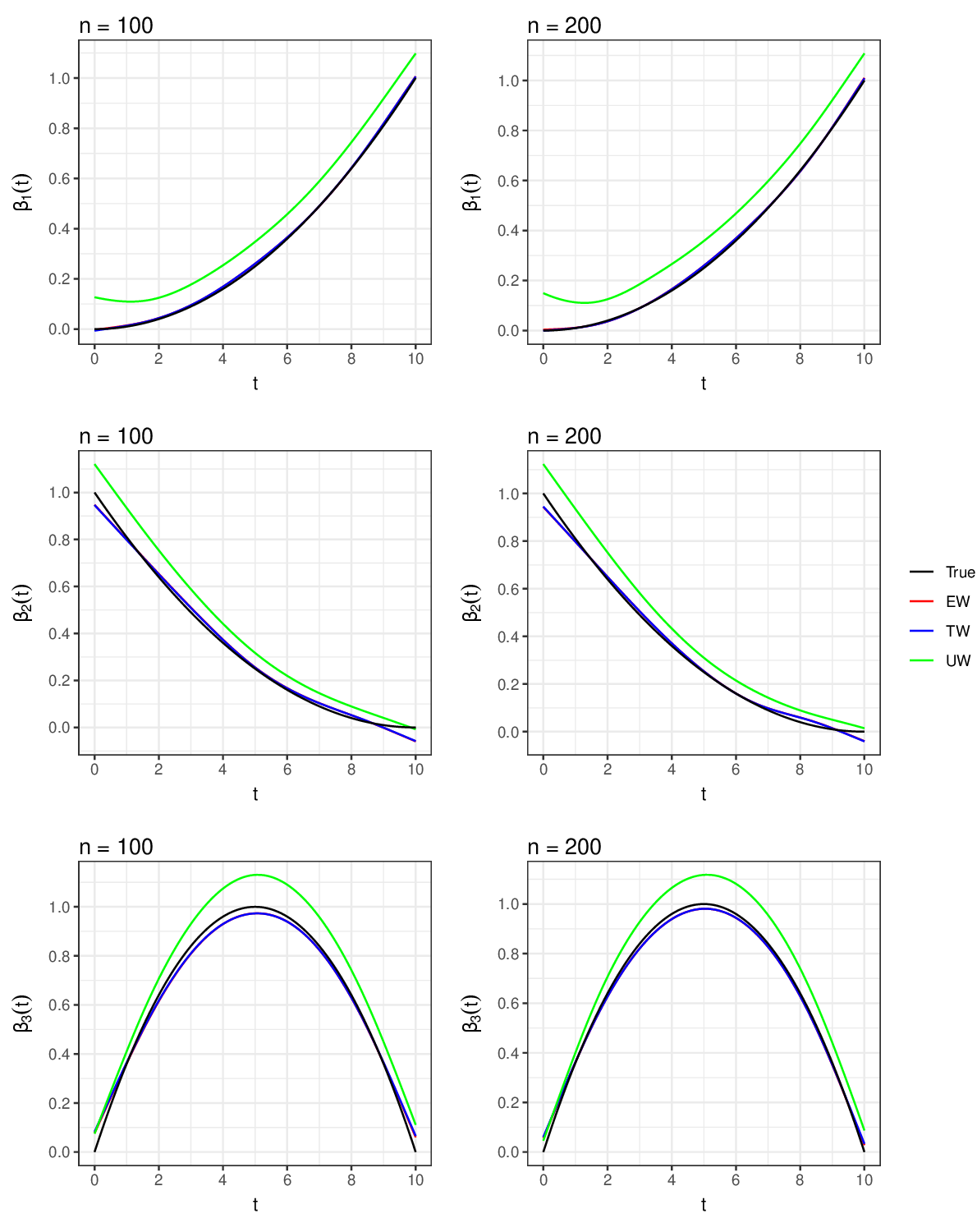}}
	\caption{Estimated varying coefficients. EW, the proposed weighted method with estimated weights; TW, the weighted method with true weights; UW, the unweighted method.}
	\label{fig:sim_beta_curve}
\end{figure}

\begin{figure}
	\centering
	{\includegraphics[width=14cm]{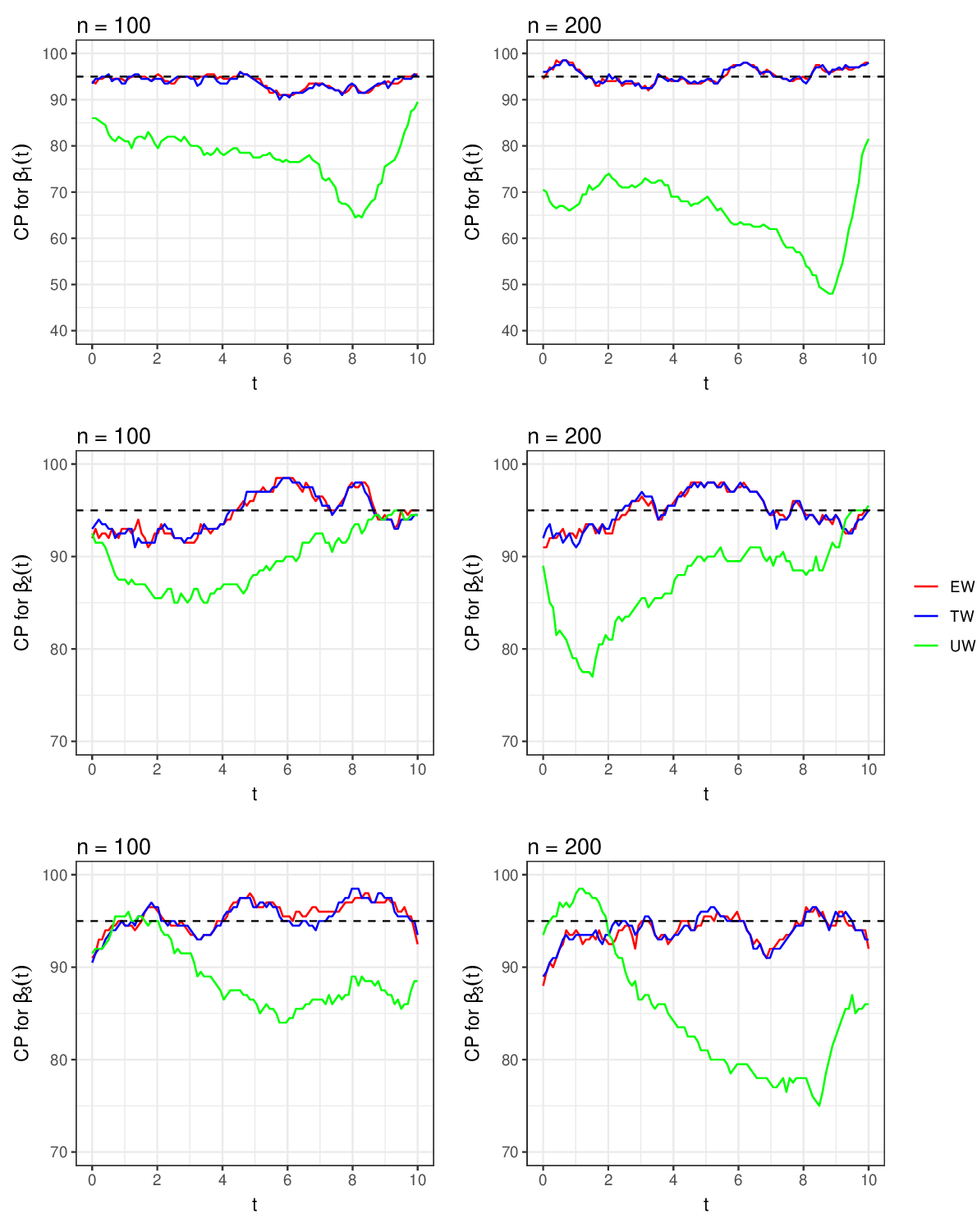}}
	\caption{Empirical coverage percentage (CP) of the 95\% pointwise confidence intervals for each varying coefficient. EW, the proposed weighted method with estimated weights; TW, the weighted method with true weights; UW, the unweighted method.}
	\label{fig:sim_beta_cp_curve}
\end{figure}

Table~\ref{tab:estimation_results_for_varying_coefficients_and_functional_principal_component_analysis} summarizes the simulation results based on 200 replicates.
Compared to the two weighted methods, the unweighted method demonstrates worse estimation performance for the varying coefficients $\mbf{\beta}(t)$, with larger estimation errors and improper overall coverage percentages of the pointwise confidence intervals. Here, the overall coverage percentage refers to the average empirical coverage probability of the confidence intervals at a dense grid points over $[0, \tau]$. 
Figures~\ref{fig:sim_beta_curve} and \ref{fig:sim_beta_cp_curve} provide a more detailed illustration through pointwise estimates and coverage percentages, further highlighting the pronounced estimation biases and poor coverage probabilities produced by the unweighted method. 
Furthermore, as shown in Table~\ref{tab:estimation_results_for_varying_coefficients_and_functional_principal_component_analysis}, the unweighted method yields larger biases and variances in estimating the covariance function, the first eigenfunction, and especially the first eigenvalue of $b_i(t)$. 
In contrast, both weighted methods achieve accurate estimation and appropriate coverage.
The results for the two weighted methods are nearly identical, indicating that the use of estimated versus true weights does not substantially affect performance.
\begin{table}[!ht]
	\caption{Estimation results for varying coefficients and functional principal component analysis. EW, the proposed weighted method with estimated weights; TW, the weighted method with true weights; UW, the unweighted method; MISE, mean integrated squared error; SD, standard deviation of the integrated squared error; CP, empirical coverage percentage of the 95\% pointwise confidence intervals, averaged over a dense grid. All numbers are based on 200 replicates.}
	{\begin{tabular}{ll*{6}{c}}
			$n$ & Par. & \multicolumn{3}{c}{MISE (SD)} & \multicolumn{3}{c}{CP} \\
			& & EW & TW & UW & EW & TW & UW \\
			100 & $\beta_1(t)$ & 0.138 (0.144) & 0.138 (0.151) & 0.223 (0.226) & 94 & 94 & 78 \\ 
			& $\beta_2(t)$ & 0.214 (0.188) & 0.213 (0.189) & 0.291 (0.395) & 95 & 95 & 90 \\ 
			& $\beta_3(t)$ & 0.199 (0.203) & 0.201 (0.199) & 0.303 (0.461) & 96 & 95 & 89 \\ 
			& $\Sigma_b(s,t)$ & 20.321 (0.294) & 20.320 (0.277) & 20.696 (1.296) &  &  &  \\ 
			& $\phi_1(t)$ & 10.693 (0.200) & 10.695 (0.199) & 10.806 (0.160) &  &  &  \\ 
			& $\theta_1$ & 0.048 (0.156) & 0.047 (0.140) & 0.275 (0.925) &  &  &  \\ \\
			200 & $\beta_1(t)$ & 0.079 (0.069) & 0.079 (0.071) & 0.200 (0.168) & 95 & 95 & 65 \\ 
			& $\beta_2(t)$ & 0.133 (0.107) & 0.134 (0.109) & 0.210 (0.338) & 95 & 95 & 87 \\ 
			& $\beta_3(t)$ & 0.134 (0.130) & 0.135 (0.129) & 0.230 (0.364) & 94 & 94 & 85 \\ 
			& $\Sigma_b(s,t)$ & 20.249 (0.110) & 20.249 (0.106) & 20.528 (0.667) &  &  &  \\ 
			& $\phi_1(t)$ & 10.703 (0.198) & 10.701 (0.200) & 10.833 (0.130) &  &  &  \\ 
			& $\theta_1$ & 0.033 (0.039) & 0.032 (0.035) & 0.136 (0.408) &  &  &  \\ 
	\end{tabular}} 
	\label{tab:estimation_results_for_varying_coefficients_and_functional_principal_component_analysis}
		
\end{table} 

We further conduct additional simulation studies with time-dependent covariates or with a non-constant baseline intensity function for the observation time process.
The proposed method continues to outperform the unweighted method and perform as well as its counterpart with true weights in both scenarios.   
Finally, we investigate how the association between the outcome and observation time processes affects the performance of the three methods by varying the coefficient of $Y_i^{obs}(t-)$ in model~\eqref{sim_PI_model} from 0 to 1.
As expected, the advantage of the weighted methods over the unweighted method increases with the strength of the association. 
When the observation time process is independent of the outcome process, all three methods achieve similar performance.
For details about these additional simulation studies, please refer to $\mathsection$S.2 of the Supplementary Material.

\section{Real data application}
\label{sec:real_data_application}
The Alzheimer's Disease Neuroimaging Initiative (ADNI) \citep{petersen2010} is a longitudinal, multi-center observational study whose primary objective is to identify and validate biomarkers for early diagnosis, monitor disease progression, and facilitate the development of effective treatments for AD. 
Since its launch in 2004, ADNI has enrolled over 2,500 participants across all phases (ADNI-1, ADNI-GO, ADNI-2, and ADNI-3), who underwent comprehensive clinical, cognitive, neuroimaging, and biological assessments at each clinical visit. 

Our analysis focuses on ADNI-1 and ADNI-2 participants with mild cognitive impairment at baseline. 
The longitudinal outcome of interest is the ADAS-Cog 13 score, which quantifies the severity of cognitive impairment across multiple domains, with higher scores indicating a worse cognitive condition. 
We include the following baseline covariates in the varying coefficient model: age, gender (male: 0 vs. female: 1), years of education, and number of ApoE-$\epsilon4$ alleles (0, 1, or 2).
In addition, we model the observation time process by a proportional intensity model that adjusts for the most recently observed ADAS-Cog 13 score (with a $\log(x+1)$ transformation for improved model fit) as well as the aforementioned baseline covariates. 
We remove all observations with missing data and all subjects with fewer than two observations. 
According to the ADNI study design, subjects are examined approximately every 6 or 12 months. 
Thus, we also remove an observation if its gap from the previous observation is smaller than a month, which is likely due to duplicated records. 
After the above data processing, there remain 807 subjects with a total of 3,558 observations.
The mean and maximum follow-up durations are 2.42 and 7.67 years, respectively, with a standard deviation of 1.26 years. 

We first fit the proportional intensity model for the observation times. 
The model fitting results suggest a significant association between the outcome and observation time processes, with the log intensity ratio for the log-transformed last observed ADAS-Cog 13 score estimated as 0.134 (standard error: 0.037, $p=0.0003$). 
We then check the adequacy of this model, including the proportional intensity assumption, the functional forms of the covariates, and the exponential link function, using numerical and graphical methods based on Schoenfeld and martingale residuals \citep{schoenfeld1982partial,therneau1990martingale}.
The global test of proportionality \citep{grambsch1994proportional} across all covariates yields a test statistic of 2.25 and a $p$-value of 0.89, indicating no evidence against the proportional intensity assumption. We further examine the functional forms of the covariates and the exponential link function based on martingale residual plots, which are presented in Figures S1 and S2 of the Supplementary Material. In all cases, the smoothed residual curves are approximately flat around 0, supporting the appropriateness of both model assumptions.

Then, we estimate the inverse-intensity weights and fit the varying coefficient model using the proposed weighted method. To avoid unstable estimation due to extremely large weights, we truncate the estimated weights at their 90\% quantile.  For comparison, we also apply the unweighted method described in $\mathsection$\ref{sec:simulation_studies} to the same dataset.
For both the weighted and unweighted methods, instead of using equally spaced knots for the B-splines as in $\mathsection$\ref{sub:estimation_of_varying_coefficients}, we place internal knots at equal quantiles of the observation times to avoid having too few observations within a time interval.  

\begin{figure}[!ht]
\centering
{\includegraphics[width=14cm]{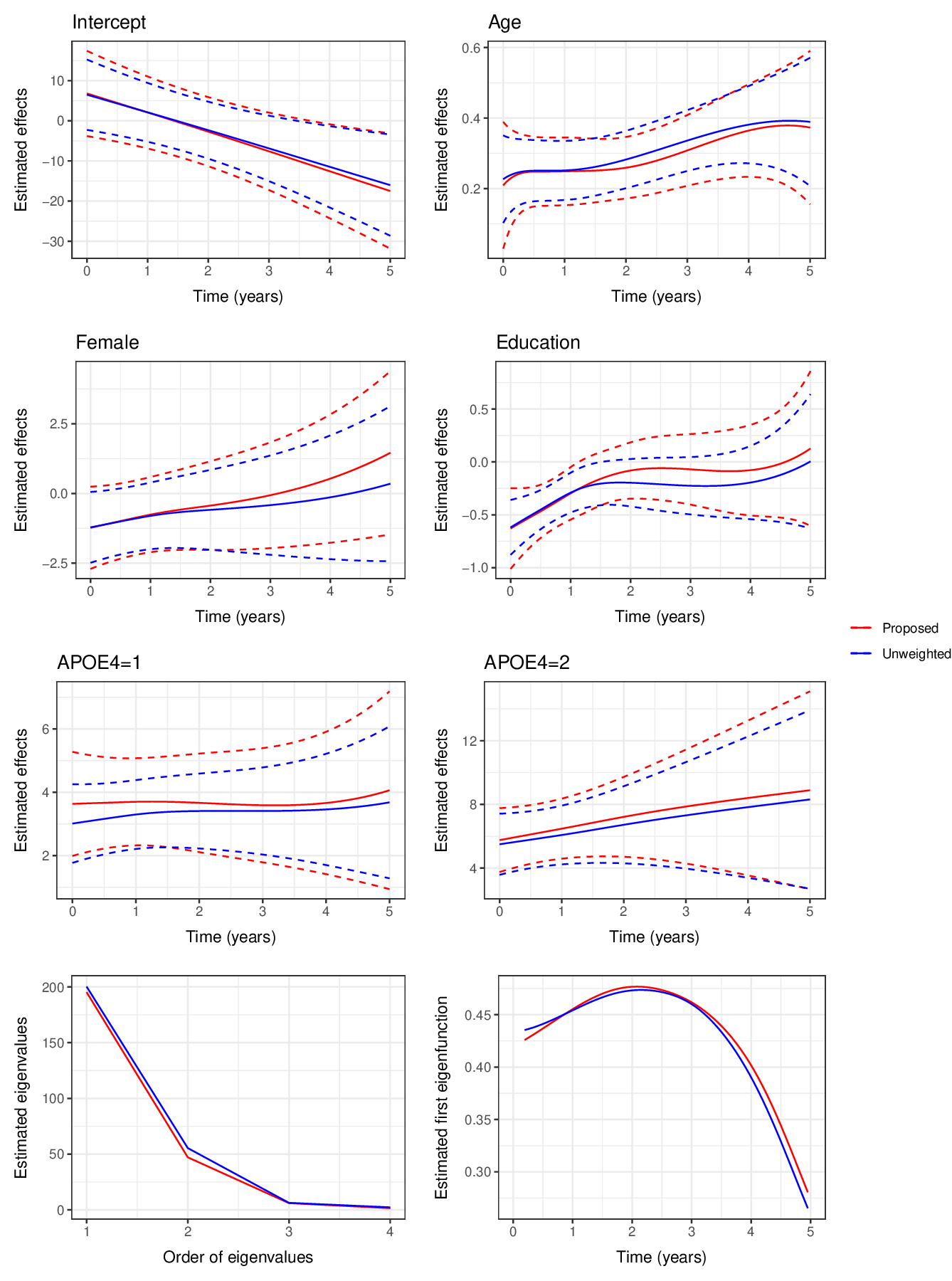}}
	\caption{Estimated varying coefficients, eigenvalues, and first eigenfunction based on the ADNI data. Dashed curves show the 95\% pointwise confidence intervals.}
	\label{fig:adni_est}
\end{figure}

The estimation results for the varying coefficients $\mbf{\beta}(t)$, eigenvalues and the first eigenfunction of $\Sigma_b$, the covariance function of $b_i(t)$, are shown in Figure~\ref{fig:adni_est}. 
Both methods suggest that older people and ApoE-$\epsilon4$ carriers have significantly worse ADAS-Cog 13 scores, while greater years of education are significantly associated with better ADAS-Cog 13 score in early stages.
These findings are consistent with those of previous studies \citep{petersen2010,kueper2018alzheimer}.
However, the unweighted method tends to overestimate the effects of age, gender, and education, and underestimate the effect of ApoE-$\epsilon4$ carrier status. 
These biases may arise because participants with worse ADAS-Cog 13 scores are more likely to attend clinic visits and remain under follow-up at later stages, thereby exerting disproportionate influence on the estimation of covariate effects. 
By incorporating inverse intensity weighting, the proposed approach mitigates these biases due to informative observation times.

\section{Discussion} \label{sec:conclusion}
In this article, we propose novel methods for estimation and inference in varying coefficient models with informative observation times. By incorporating inverse intensity weighting into sieve estimation, we correct biases induced by informative sampling and obtain consistent, asymptotically normal estimators for the coefficient functions. Through extensive simulation studies and an application to the ADNI study, we show that the proposed approach substantially outperforms conventional unweighted analyses, yielding smaller estimation biases and appropriate coverage probabilities of pointwise confidence intervals.

The current framework can be naturally extended to multivariate settings. For example, in addition to the ADAS-Cog 13 score, the ADNI study also collects longitudinal measurements on several other cognitive biomarkers, including the Rey Auditory Verbal Learning Test (RAVLT) immediate recall, RAVLT learning, and the Mini-Mental State Examination. It is clinically meaningful to analyze these outcomes jointly, accounting for their dependence. One could consider a multivariate varying coefficient model \citep{zhu2012multivariate} for these longitudinal outcomes and incorporate the same inverse intensity weighting strategy used here into the estimation procedure to correct for biases induced by informative observation times. Another important direction is to extend the current framework to generalized varying coefficient models \citep{csenturk2008generalized} to handle non-continuous longitudinal outcomes, such as counts and categorical data. Again, inverse intensity weighting can be incorporated into existing estimation frameworks to address informative observation times.

Finally, our current work assumes that the follow-up time $C_i$ is independent of the outcome process. This assumption may be violated in practice. For instance, longitudinal trajectories may be truncated by outcome-related dropout or by terminal events such as death, whose occurrence may be informative about the outcome process itself. In such scenarios, $C_i$ is associated with the outcome process, and our proposed estimators are generally biased. To address this issue, one could specify an additional Cox-type model for $C_i$, adjusting for the past observed outcome and covariate history, and then incorporate the inverse probability of remaining under follow-up at each observation time into the current weighting scheme.

\section*{Supplementary material}
\label{SM}
We provide the proofs of Theorems~\ref{thm-slopeconsistency} and \ref{thm-slopeasymnorm} in $\mathsection$\ref{sec:theory}, as well as additional numerical results in $\mathsection$\ref{sec:simulation_studies} and $\mathsection$\ref{sec:real_data_application}.


\bibliographystyle{apalike}
\bibliography{paper-ref}

@book{andersen1993statistical,
  title={{Statistical Models Based on Counting Processes}},
  author={Andersen, Per K and Borgan, Ornulf and Gill, Richard D and Keiding, Niels},
  year={1993},
  publisher={New York: Springer}
}

@book{vaart1996weak,
  title={Weak Convergence and Empirical Processes},
  author={van der Vaart, A. W. and Wellner, J. A.},
  year={1996},
  publisher={New York: Springer}
}

@book{gu2013smoothing,
  title={Smoothing Spline ANOVA Models},
  author={Gu, Chong},
  year={2013},
  publisher={New York: Springer}
}

@article{yao2005functional,
  title={Functional data analysis for sparse longitudinal data},
  author={Yao, Fang and M{\"u}ller, Hans-Georg and Wang, Jane-Ling},
  journal={Journal of the American Statistical Association},
  volume={100},
  number={470},
  pages={577--590},
  year={2005},
  publisher={Taylor \& Francis}
}

@Manual{zhou2024fdapace,
    title = {{fdapace: Functional Data Analysis and Empirical
      Dynamics}},
    author = {Yidong Zhou and Han Chen and Su I Iao and Poorbita Kundu
      and Hang Zhou and Satarupa Bhattacharjee and Cody Carroll and
      Yaqing Chen and Xiongtao Dai and Jianing Fan and Alvaro Gajardo
      and Pantelis Z. Hadjipantelis and Kyunghee Han and Hao Ji and
      Changbo Zhu and Hans-Georg Müller and Jane-Ling Wang},
    year = {2024},
    note = {R package version 0.6.0},
  }

@article{petersen2010,
author = {Petersen, R. C. and Aisen, P. S. and Beckett, L. A. and Donohue, M. C. and Gamst, A. C. and Harvey, D. J. and Jack, C. R. and Jagust, W. J. and Shaw, L. M. and Toga, A. W. and Trojanowski, J. Q. and Weiner, M. W.},
title = {Alzheimer's Disease Neuroimaging Initiative (ADNI)},
journal = {Neurology},
volume = {74},
number = {3},
pages = {201-209},
year = {2010}
}

@article{kueper2018alzheimer,
  title={The {Alzheimer’s} disease assessment scale--cognitive subscale (ADAS-Cog): modifications and responsiveness in pre-dementia populations. a narrative review},
  author={Kueper, Jacqueline K and Speechley, Mark and Montero-Odasso, Manuel},
  journal={Journal of Alzheimer’s Disease},
  volume={63},
  number={2},
  pages={423--444},
  year={2018},
  publisher={SAGE Publications Sage UK: London, England}
}

@article{zhu2012multivariate,
  title={Multivariate varying coefficient model for functional responses},
  author={Zhu, Hongtu and Li, Runze and Kong, Linglong},
  journal={The Annals of Statistics},
  volume={40},
  number={5},
  pages={2634--2666},
  year={2012}
}

@article{zhang2013time,
  title={Time-varying additive models for longitudinal data},
  author={Zhang, Xiaoke and Park, Byeong U and Wang, Jane-Ling},
  journal={Journal of the American Statistical Association},
  volume={108},
  number={503},
  pages={983--998},
  year={2013},
  publisher={Taylor \& Francis}
}

@article{fan2007analysis,
  title={Analysis of longitudinal data with semiparametric estimation of covariance function},
  author={Fan, Jianqing and Huang, Tao and Li, Runze},
  journal={Journal of the American Statistical Association},
  volume={102},
  number={478},
  pages={632--641},
  year={2007},
  publisher={Taylor \& Francis}
}

@article{weaver2023functional,
  title={Functional data analysis for longitudinal data with informative observation times},
  author={Weaver, Caleb and Xiao, Luo and Lu, Wenbin},
  journal={Biometrics},
  volume={79},
  number={2},
  pages={722--733},
  year={2023},
  publisher={Wiley Online Library}
}

@article{lin2004analysis,
  title={Analysis of longitudinal data with irregular, outcome-dependent follow-up},
  author={Lin, Haiqun and Scharfstein, Daniel O and Rosenheck, Robert A},
  journal={Journal of the Royal Statistical Society Series B: Statistical Methodology},
  volume={66},
  number={3},
  pages={791--813},
  year={2004},
  publisher={Oxford University Press}
}

@article{xu2024bias,
  title={Bias-correction and test for mark-point dependence with replicated marked point processes},
  author={Xu, Ganggang and Zhang, Jingfei and Li, Yehua and Guan, Yongtao},
  journal={Journal of the American Statistical Association},
  volume={119},
  number={545},
  pages={217--231},
  year={2024},
  publisher={Taylor \& Francis}
}

@article{stone1985,
  title={Additive regression and other nonparametric models},
  author={Stone, Charles J},
  journal={The Annals of Statistics},
  volume={13},
  number={2},
  pages={689--705},
  year={1985},
  publisher={Institute of Mathematical Statistics}
}

@article{li2010uniform,
  title={Uniform convergence rates for nonparametric regression and principal component analysis in functional/longitudinal data},
  author={Li, Yehua and Hsing, Tailen},
  journal={The Annals of Statistics},
  volume={38},
  number={6},
  pages={3321--3351},
  year={2010},
  publisher={Institute of Mathematical Statistics}
}

@article{zhang2016,
  title={From sparse to dense functional data and beyond},
  author={Zhang, Xiaoke and Wang, Jane-Ling},
  journal={The Annals of Statistics},
  volume={44},
  number={5},
  pages={2281--2321},
  year={2016},
  publisher={Institute of Mathematical Statistics}
}

@article{cox1972,
  title={Regression models and life-tables},
  author={Cox, David R},
  journal={Journal of the Royal Statistical Society: Series B (Methodological)},
  volume={34},
  number={2},
  pages={187--202},
  year={1972},
  publisher={Wiley Online Library}
}

@book{tsybakov2009,
  title={Introduction to Nonparametric estimators},
  author={Tsybakov, Alexandre B},
  year={2009},
  publisher={New York: Springer}
}

@article{zhang2015varying,
  title={Varying-coefficient additive models for functional data},
  author={Zhang, Xiaoke and Wang, Jane-Ling},
  journal={Biometrika},
  volume={102},
  number={1},
  pages={15--32},
  year={2015},
  publisher={Oxford University Press}
}

@article{stone1982,
  title={Optimal global rates of convergence for nonparametric regression},
  author={Stone, Charles J},
  journal={The Annals of Statistics},
  volume={10},
  number={4},
  pages={1040--1053},
  year={1982},
  publisher={JSTOR}
}

@article{sang2025functional,
  title={Functional principal component analysis with informative observation times},
  author={Sang, Peijun and Kong, Dehan and Yang, Shu},
  journal={Biometrika},
  volume={112},
  number={1},
  pages={asae055},
  year={2025},
  publisher={Oxford University Press}
}

@article{zhou2025theory,
  title={Theory of functional principal component analysis for discretely observed data},
  author={Zhou, Hang and Wei, Dongyi and Yao, Fang},
  journal={The Annals of Statistics},
  volume={53},
  number={5},
  pages={2103--2127},
  year={2025},
  publisher={Institute of Mathematical Statistics}
}

@article{dette2024statistical,
  title={Statistical inference for function-on-function linear regression},
  author={Dette, Holger and Tang, Jiajun},
  journal={Bernoulli},
  volume={30},
  number={1},
  pages={304--331},
  year={2024},
  publisher={Bernoulli Society for Mathematical Statistics and Probability}
}

@article{xie2023scalable,
  title={Scalable inference in functional linear regression with streaming data},
  author={Xie, Jinhan and Shi, Enze and Sang, Peijun and Shang, Zuofeng and Jiang, Bei and Kong, Linglong},
  journal={The Annals of Statistcs},
  volume={forthcoming},
  year={2025}
}

@article{huang2004polynomial,
 author = {Jianhua Z. Huang and Colin O. Wu and Lan Zhou},
 journal = {Statistica Sinica},
 number = {3},
 pages = {763--788},
 publisher = {Institute of Statistical Science, Academia Sinica},
 title = {POLYNOMIAL SPLINE ESTIMATION AND INFERENCE FOR VARYING COEFFICIENT MODELS WITH LONGITUDINAL DATA},
 urldate = {2025-12-12},
 volume = {14},
 year = {2004}
}

@article{muller1994hazard,
  title={Hazard rate estimation under random censoring with varying kernels and bandwidths},
  author={M{\"u}ller, Hans-Georg and Wang, Jane-Ling},
  journal={Biometrics},
  volume={50},
  number={1},
  pages={61--76},
  year={1994},
  publisher={JSTOR}
}

@article{grambsch1994proportional,
  title={Proportional hazards tests and diagnostics based on weighted residuals},
  author={Grambsch, Patricia M and Therneau, Terry M},
  journal={Biometrika},
  volume={81},
  number={3},
  pages={515--526},
  year={1994},
  publisher={Oxford University Press}
}

@article{schoenfeld1982partial,
  title={Partial residuals for the proportional hazards regression model},
  author={Schoenfeld, David},
  journal={Biometrika},
  volume={69},
  number={1},
  pages={239--241},
  year={1982},
  publisher={Oxford University Press}
}

@article{therneau1990martingale,
  title={Martingale-based residuals for survival models},
  author={Therneau, Terry M and Grambsch, Patricia M and Fleming, Thomas R},
  journal={Biometrika},
  volume={77},
  number={1},
  pages={147--160},
  year={1990},
  publisher={Oxford University Press}
}

@article{pullenayegum2016longitudinal,
  title={Longitudinal data subject to irregular observation: {A} review of methods with a focus on visit processes, assumptions, and study design},
  author={Pullenayegum, Eleanor M and Lim, Lily SH},
  journal={Statistical Methods in Medical Research},
  volume={25},
  number={6},
  pages={2992--3014},
  year={2016},
  publisher={SAGE Publications Sage UK: London, England}
}

@article{zeng2005asymptotic,
  title={Asymptotic results for maximum likelihood estimators in joint analysis of repeated measurements and survival time},
  author={Zeng, Donglin and Cai, Jianwen}, 
  journal={The Annals of Statistics},
  volume={33},
  number={5},
  pages={2132-2163},
  year={2005}
}

@article{csenturk2008generalized,
  title={Generalized varying coefficient models for longitudinal data},
  author={{\c{S}}ent{\"u}rk, Damla and M{\"u}ller, Hans-Georg},
  journal={Biometrika},
  volume={95},
  number={3},
  pages={653--666},
  year={2008},
  publisher={Oxford University Press}
}

@article{Jack2013,
    author = {Jack, Clifford R, Jr and Knopman, David S and Jagust, William J and Petersen, Ronald C and Weiner, Michael W and Aisen, Paul S and Shaw, Leslie M and Vemuri, Prashanthi and Wiste, Heather J and Weigand, Stephen D and Lesnick, Timothy G and Pankratz, Vernon S and Donohue, Michael C and Trojanowski, John Q},
    date = {2013/02/01},
    doi = {10.1016/S1474-4422(12)70291-0},
    isbn = {1474-4422},
    journal = {The Lancet Neurology},
    journal1 = {The Lancet Neurology},
    month = {2026/01/03},
    number = {2},
    pages = {207--216},
    publisher = {Elsevier},
    title = {Tracking pathophysiological processes in Alzheimer's disease: an updated hypothetical model of dynamic biomarkers},
    type = {doi: 10.1016/S1474-4422(12)70291-0},
    volume = {12},
    year = {2013},
    year1 = {2013}
}

@article{Huang2002,
    author = {Huang, Jianhua Z. and Wu, Colin O. and Zhou, Lan},
    title = {Varying‐coefficient models and basis function approximations for the analysis of repeated measurements},
    journal = {Biometrika},
    volume = {89},
    number = {1},
    pages = {111-128},
    year = {2002},
    month = {03},
    issn = {0006-3444},
    doi = {10.1093/biomet/89.1.111},
    eprint = {https://academic.oup.com/biomet/article-pdf/89/1/111/591363/890111.pdf},
}

@article{Hoover1998,
    author = {Hoover, Donald R. and Rice, John A. and Ww, Colin O. and Yang, Li-Ping},
    title = {Nonparametric smoothing estimates of time-varying coefficient models with longitudinal data},
    journal = {Biometrika},
    volume = {85},
    number = {4},
    pages = {809-822},
    year = {1998},
    month = {12},
    issn = {0006-3444},
    doi = {10.1093/biomet/85.4.809},
    eprint = {https://academic.oup.com/biomet/article-pdf/85/4/809/698554/85-4-809.pdf},
}

@article{Breslow,
    Author = {Breslow, N. E.},
    Journal = {J. R. Statist. Soc. {\rm B}},
    Pages = {216--217},
    Title = {{Discussion of the paper by D. R. Cox.}},
    Volume = {34},
    Year = {1972}
}

@article{Aalen1978,
author = {Odd Aalen},
title = {{Nonparametric Inference for a Family of Counting Processes}},
volume = {6},
journal = {The Annals of Statistics},
number = {4},
publisher = {Institute of Mathematical Statistics},
pages = {701 -- 726},
year = {1978}
}

@article{claeskens2009asymptotic,
  title={Asymptotic properties of penalized spline estimators},
  author={Claeskens, Gerda and Krivobokova, Tatyana and Opsomer, Jean D},
  journal={Biometrika},
  volume={96},
  number={3},
  pages={529--544},
  year={2009},
  publisher={Oxford University Press}
}

\end{document}